\documentclass{elsart}

\usepackage{graphicx}
\usepackage{dcolumn}
\usepackage{bm}

\usepackage{amssymb}

\begin{document}

\begin{frontmatter}



\title{\large Gradient Pattern Analysis of Cosmic Structure Formation: 
Norm and Phase Statistics}


\author{A.P.A. Andrade\thanksref{1}}, \author{ A.L. B. Ribeiro\thanksref{2}}
\& \author{R.R. Rosa\thanksref{3}}

\thanks[1]{apaula@uesc.br} 
\thanks[2]{albr@uesc.br}
\address{ 1,2 - Laborat\'orio de Astrof\'isica Te\'orica e Observacional,
  Departamento de Ci\^encias Exatas e Tecnol\'ogicas, Universidade 
Estadual de Santa Cruz, Rodovia Ilh\'eus-Itabuna km 16,
  45662.000, Ilh\'eus, Brasil.}

\thanks[3]{reinaldo@lac.inpe.br}
\address{3 - Laborat\'orio Associado de Computa\c c\~ao e Matem\'atica 
Aplicada, Instituto Nacional de Pesquisas Espaciais, 12201-970, S\~ao Jos\'e 
dos Campos, Brasil.}

\begin{abstract}
This paper presents the preliminary results of the 
characterization of pattern evolution in the process of cosmic structure
formation. We are applying on N-body cosmological simulations data the
technique proposed by Rosa, Sharma \& Valdivia (1999) and Ramos et al. 
(2000) to estimate the time behavior of asymmetries in the gradient field.
The gradient pattern analysis is a well tested tool, used to build asymmetrical fragmentation 
parameters estimated over a gradient field of an image matrix able to quantify a complexity measure of 
nonlinear extended systems. In this investigation we work with the high resolution 
cosmological data simulated by the Virgo consortium, in different time steps,
in order to obtain a diagnostic of the spatio-temporal
disorder in the matter density field. We perform the calculations of the gradient vectors statistics, such as mean, variance, skewness, 
kurtosis, and correlations on the gradient field. Our main goal is to determine 
different dynamical regimes through the analysis of complex patterns arising from the 
evolutionary process of structure formation. The results show that
the gradient pattern technique, specially the statistical analysis of second and third 
gradient moment, may represent a very useful tool to describe the matter
clustering in the Universe.
\end{abstract}

\end{frontmatter}

\section{Introduction}
\label{sec:level1}
Understanding how the spatio-temporal organization of matter happens in
the Universe is an important task in modern science. The standard model
for cosmic structure formation considers the existence of small inhomogeneities in the matter density field which grew up, in an isotropically expanding universe, trough gravitational
clustering. As the fluctuation become intensified, they start to 
interact which other performing a highly nonlinear evolution during the matter
clustering, leading to a cosmic matter distribution settles into an
anisotropic patterns.  In this dynamical 
regime, phase coupling can not be avoided, and we can expect a nontrivial asymmetrical 
pattern arising as the system evolves to a highly inhomogeneous field, being composed by galaxies accumulated in walls, filaments and dense compact
clusters, interlocked by large voids regions, characterizing a {\it foamlike 
appearance} in the cosmic matter distribution (Weygaert, 2002). 
Over the nonlinear regime, N-body simulations are used to follow the 
later evolution of the matter content in the Universe, which is dominated by 
weakly interacting particles, being the gravity the main interacting 
force, but also considering corrections factors for dissipational physical effects and
hydrodynamical processes (Springel et al., 2005). 

The main tools used in cosmology to investigate the evolutionary process of structure formation and compare
theoretical models with observations are statistical (Martinez \& Saars,
2002). Usually, they are based on a spatial statistics (global mean-field) or
a local (nearest-neighbor) analysis. The most popular 
tools used to describe the large scale structures are the spatial
statistics such as the Fourier analysis and the N-point correlation
function (e.g. Bardeen et al., 1986). However, other clustering measure, like
the local statistics, are being extensively used to provide complementary 
information to the spatial statistics. They include 
the topological investigation of the density field, such as the nearest neighbor
statistics, the genus and the Minkowski functional (e.g.  Mecke, Buchert \&
Wagner, 1994); the stochastic geometry analysis, which provides 
a geometrical model of the cellular distribution of matter in the Universe
(Bernardeau \& Weygaert, 1996; Weygaert, 2002); and the phase mapping analysis used to describe phase
patterns in Fourier space (Chiang, Coles \& Naselsky, 2002). All these tools
complement each other in the investigation of cosmic
structure formation and evolution. The purpose of this paper, is to test 
the ability of the {\it gradient pattern analysis} (GPA)
technique to help the investigation of the spatial-temporal organization of matter in the
Universe. We hope this new approach may represent a simple and useful
tool to investigate the cosmic structure clustering, as well to compare
structural theoretical predictions with observations, since the GPA is a very sensitive
technique and, in contrast with other nearest-neighbor methods, does not require large number of
points. Actually, GPA has been used to characterize spatial structures from the point of view of their irregularities meanwhile most the techniques are based on the presence of regularities. 

The GPA technique was proposed by Rosa, Sharma \& Valdivia (1999), in order to characterize and quantify pattern
formation and evolution in complex systems. The measures obtained by GPA is based on the 
spatio-temporal local density amplitude fluctuations on a structure
represented as a dynamical gradient pattern, which may characterize nonlinear
emergence and evolution of ordered structures from a random initial condition. 
This technique has being successfully applied to several nonlinear problems in physics, chemistry and biology 
fields, being able to characterize symmetry breaks, extended pattern evolution,
as well the spatio-temporal relaxation in complex systems (Rosa et al., 2003
and references there in). At this moment, the GPA technique is also being applying on
topological feature investigation in the cosmic microwave background maps
(Rosa et al., 2006). In this work, we give a particular attention to the
investigation of the second and third gradient moments to describe the pattern
formation and evolution of cosmic structure formation in a two-dimensional mesh. 
For this purpose, we perform the statistical analysis of norms and phases of
the gradient vectors for the particle density field in N-Body simulations. 
We estimate mean, variance, skewness, kurtosis and correlations between the
gradient vectors. In order to explore symmetries break, we have
extended the calculation for both asymmetric and the whole gradient 
field.

\section{The Gradient Pattern Analysis}
\label{sec:level1}

If we consider spatial extended systems in two dimensions (x,y) 
their
amplitude distribution can be represented as an envelope A(x,y),
which can be approximated by a square matrix of amplitude ${\bf A}={\it
  \{a_ {m,m} \} }$, in which the two dimensions, {\it x} and {\it y}, 
are discretized into M x M
pixels. Thus, a dynamical sequence of N matrices is 
related to a temporal
evolution of a visualized envelope A(x,y,t). The spatial 
fluctuation
of the global pattern A(x,y) can be represented by its gradient 
vector
field $ G_t = \nabla[A(x,y)]_t $. The local spatial fluctuations, 
between a
pair of neighbor cells, of the global pattern is characterized by its 
gradient
vector at corresponding mesh-points in two dimensional space. In this
representation, the relative values between cells $(\Delta A \equiv 
|A(i,j)-A(i+1,j+1)|)$
are dynamically relevant, rather than the pixels absolute values. 
Indeed, in a
gradient field such relatives values, $\Delta A $, can be characterized 
by
each local vector norm $(r_{i,j})$ and phase $(\phi_{i,j})$. Thus a 
given
scalar field can be represented by a gradient field for the local 
amplitude
fluctuations, and its gradient pattern can be represented by a pair of
matrices, one for norms and other for phases. Indeed, a natural 
representation
is by means of a complex matrix, where each element corresponds to a
respective complex number $z_{i,j}$ representing each vector from the
gradient pattern. Thus, a given matricial scalar field can be 
represented as a composition of four gradient moments (Rosa  et al., 2003: 
 first order, $ g_1$, is the global representation of the vectors 
distribution;
 second order, $g_2$, is the global representation of norms; the third, 
$g_3$, is the global representation of phases; 
 and fourth order, $g_4$, the global complex representation of the 
gradient pattern. In this work, we are focusing on the 
second and third gradient moments.

\subsection{The First Gradient Moment}
A possible information that we can extract from the distribution of 
the
gradient vector field is about symmetry breaking between pairs of 
vectors. Let
us consider two vectors $V_i$ and $V_j$, with $i \neq j$ belonging to 
the 
$G_t$ field. We could say that  $V_i$ and $V_j$ are vectorially 
symmetric if
$V_i = -V_j$, so that $R_{i,j} = V_i -V_j =0$. In computer terms, we 
could
say that the resulting vector, $R_{i,j}$, is null within a tolerance
$\epsilon$. The $\epsilon$ value would be established according to 
distribution of the
gradient vectors, e.g., equal to one tenth of: (a) standard deviation 
of
magnitudes, and (b) misalignment between them (Rosa, Sharma \& 
Valdivia, 1998).
If we remove every pair of symmetric vectors from the $G_t$ field, we 
will be
generating a field with {\it L} vectors, all of them vectorially 
asymmetric. 
The parameter {\it L} would be a first measure of asymmetries in the 
gradient
field. For the asymmetrical vector set, we could also try to perform a
statistical analysis of norms and phases. In Figure 1 we present the 
complete schematic representation of the Gradient
Pattern Analysis of a bi-dimensional scalar matricial field. It is also possible to extend the same approach for a data set in 
three dimensions (x,y,z), building an amplitude matrix A(x,y,z,t), similarly ${\bf A}={\it \{a_ {m,m,m} \}} $ and a three-dimensional gradient field $ G_t = \nabla[A(x,y,z)]_t $.

\begin{figure}
\begin{center}
\includegraphics[width=5.0in, height=3.99in]{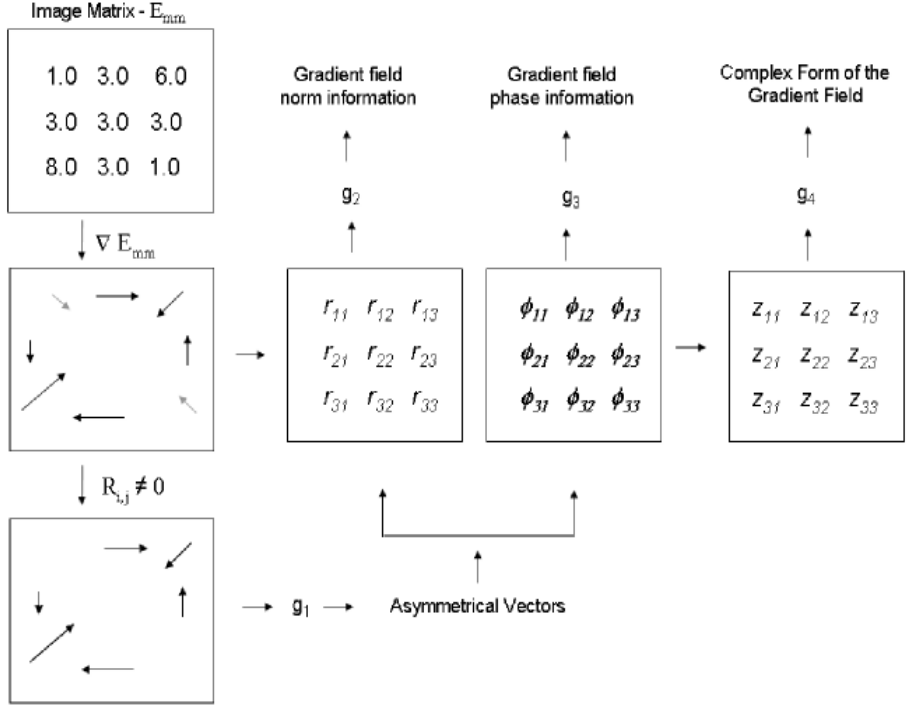}
\caption{\label{fig:epsart} GPA Summry: building the gradient moments 
from an image scale matrix}
\end{center}
\end{figure}

\subsection{The Second and Third Gradient Moment}

For the second and third moments, there is no formal computational 
operation to calculate them in literature. However, to extract basic information 
about
amplitudes and orientation distribution of the gradient field  vectors, it is possible
to perform a basic statistical analysis between vectors by the norms $(r_{i,j})$ 
and phases $(\phi_{i,j})$ matrix. This statistical analysis could include:

\begin{itemize}
\item mean: $\left < r_{i,j} \right >$ \ and \ $\left < \phi_{i,j} 
\right >$;\\
\item  variance: $ \left <(r_{i,j} - \left < r_{i,j} \right >)^2 \right 
>$ and $\left < ( \phi_{i,j}- \left < \phi_{i,j} \right >)^2 \right 
>$;\\
\item skewness: $\frac {\left < (r_{i,j} - \left < r_{i,j} \right >)^3 
\right >} {\left <
(r_{i,j} - \left < r_{i,j} \right >)^2 \right >^\frac{3}{2}} $ \ and \ 
$
\frac {\left < (\phi_{i,j} - \left < \phi_{i,j} \right >)^3 \right >} 
{\left <
(\phi_{i,j} - \left < \phi_{i,j} \right >)^2 \right >^\frac{3}{2}} $ 
;\\
\item kurtosis: $
\frac {\left <
(r_{i,j} - \left < r_{i,j} \right >)^4 \right >} {\left <
(r_{i,j} - \left < r_{i,j} \right >)^2 \right >^2} $ \ and \ $
\frac {\left <
(\phi_{i,j} - \left < \phi_{i,j} \right >)^4 \right >} {\left <
(\phi_{i,j} - \left < \phi_{i,j} \right >)^2 \right >^2} $;\\
\item complex correlation: $ \left < z_{i,j}z_{l,m} \right >$.
\end{itemize}

\section{Data Base and Analysis}
\label{sec:level1}

For the first check of GPA ability in the process of cosmic structure formation, we 
have chosen to perform the analysis only in a two-dimensional 
mesh estimated over the public domain high resolution N-body cosmological data 
simulated by the {\it Virgo Consortium (http://www.mpa-garching.mpg.de/Virgo/)}  
for a $\Lambda$-CDM Universe (Jenkins et al., 1998). To assure a good 
statistical assembly, we have computed 150 bi-dimensional amplitude matrix for the density field, {\it $E_{mn}$ }, in   
seventeen time step (redshift, z) on interval $(0 \le z \le 10)$, computed in a mesh grid of $(180 \times 180)$,  
estimated over a simulated volume of $(120 \times 120 \times 
10)h^{-3}Mpc^3$. This discretized mesh enable us to deal with local amplitude 
fluctuations with resolution of $\sim 0.7 Mpc~h^{-1}$, scale in which we can
map organizing substructures inside galaxy clusters. In Figure 2, we present 
three of this bi-dimensional slices of the simulated volume for three redshift  
values:  z = 10, 3 and 0. Also in Figure 2, we show the frames of their 
respective intensity contours plot. The matter organization in clusters,  
filaments and voids is clearly seen in this frames. 
Remembering that higher redshift represents earlier epochs, while  
the redshift zero is the present time. 
 
\begin{figure}
\begin{center}
\includegraphics[width=6.0in, height=2.7in]{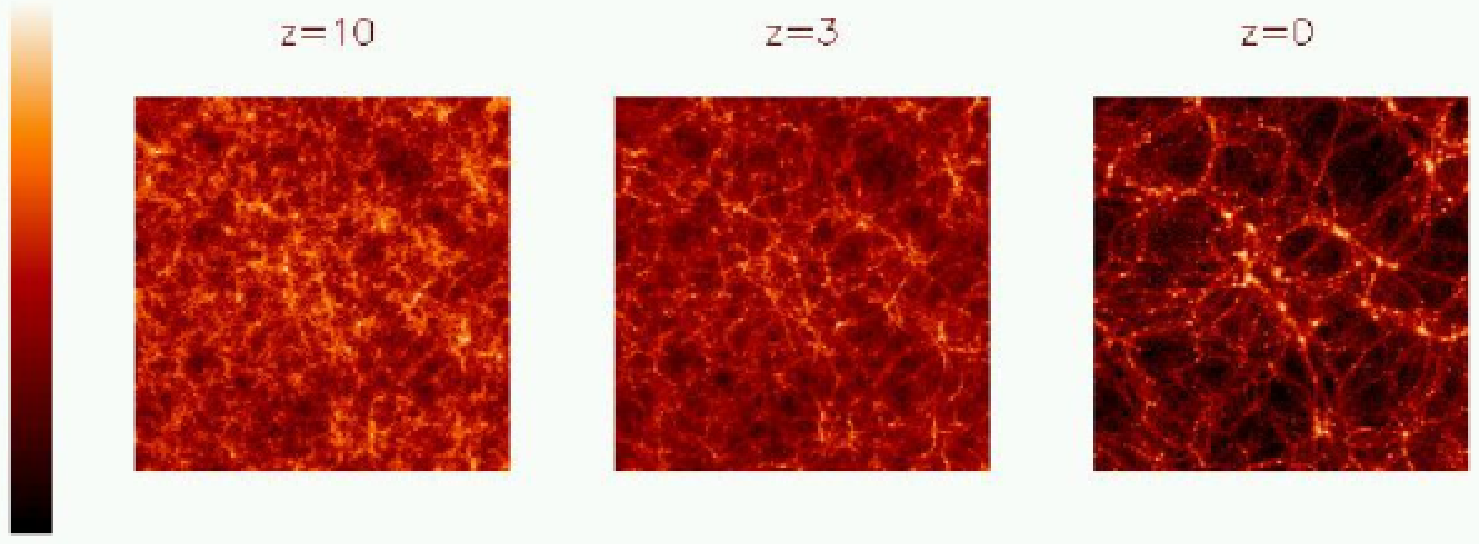}
\includegraphics[width=5.7in, height=2.1in]{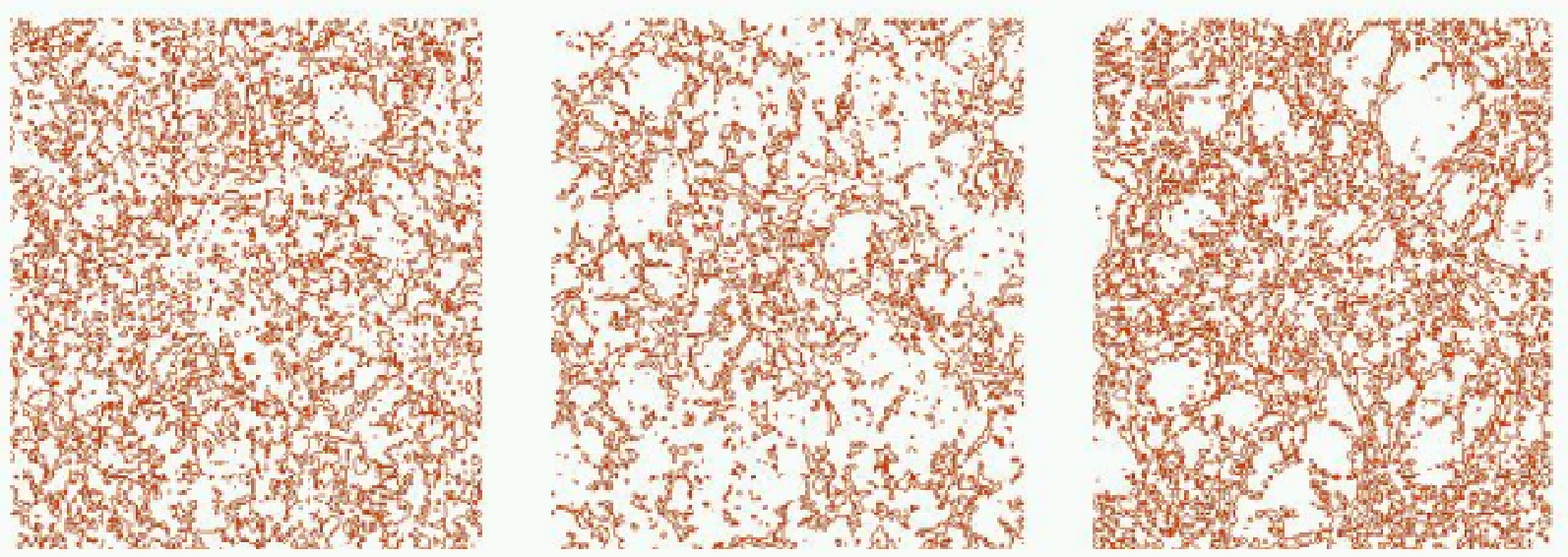}
\caption{\label{fig:epsart} Three slices of a tri-dimensional N-body
  simulations for a $\Lambda$-CDM universe on redshifts: 10, 3 and 0. The frames on the bottom are their
  respective intensity contours plot.}
\end{center}
\end{figure}

In order to test the efficiency of the GPA technique to
characterize pattern evolution in process of structure formation, 
we have performed a detailed statistical analysis between norms and phases, 
the second and third gradient moments of the gradient field. We have worked with
every 150 volume slices for each one of the seventeen 
redshift intervals simulated. In the final analysis we have performed the mean 
value between the 150 slices to compute the statistical moment. We have
considered independently the analysis in two cases: first, for all the 
vectors in the gradient field and second, only for the asymmetrical vectors
estimated in the gradient field, as discussed in section 2.1. We also have
performed a simulation of eighteen set of 150 random matrices, 
with mean amplitude in the same interval as the N-body data, in order to 
compare the statistical analysis of the norms and phase vectors with a random
distribution. The main results obtained are illustrated in Figures 3, 4 and 5. In Figure 3 we present
the histogram of norms and phases of the gradient field. The statistical
moments (mean, variance, skewness and kurtosis) of the norms and phases are
plotted in Figures 4 and 5, respectively.

\begin{figure}
\includegraphics[width=2.8in, height=2.8in]{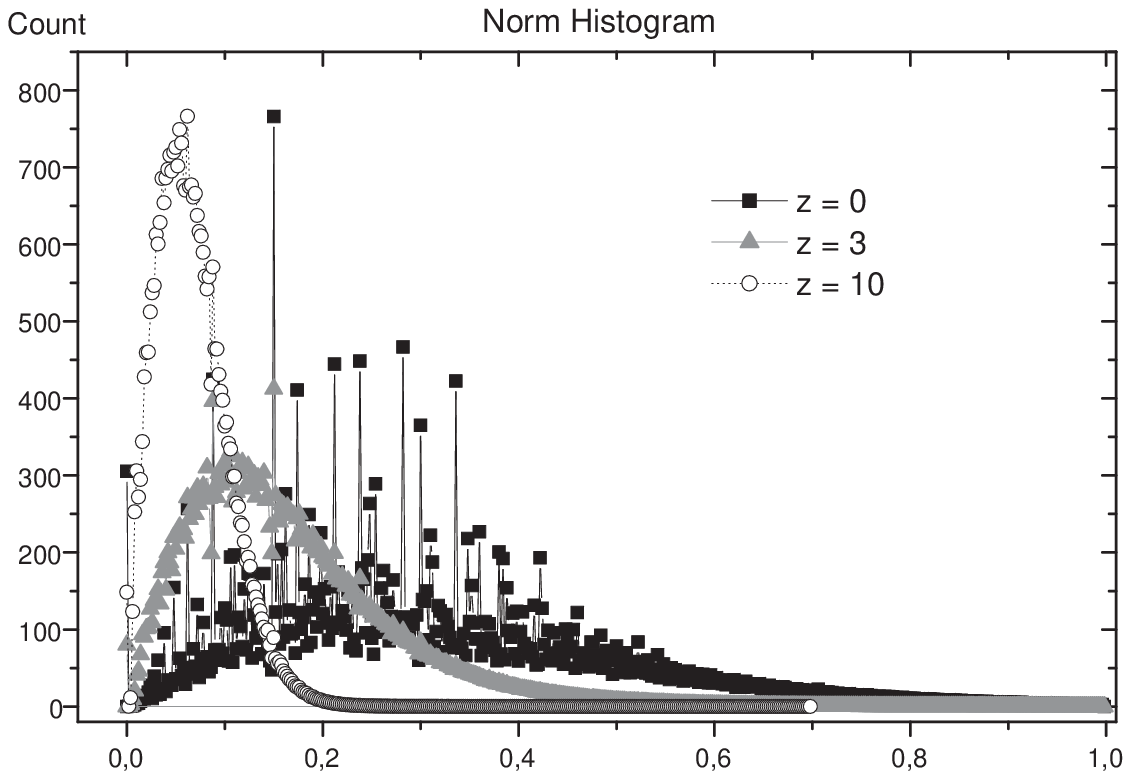}
\includegraphics[width=2.8in, height=2.8in]{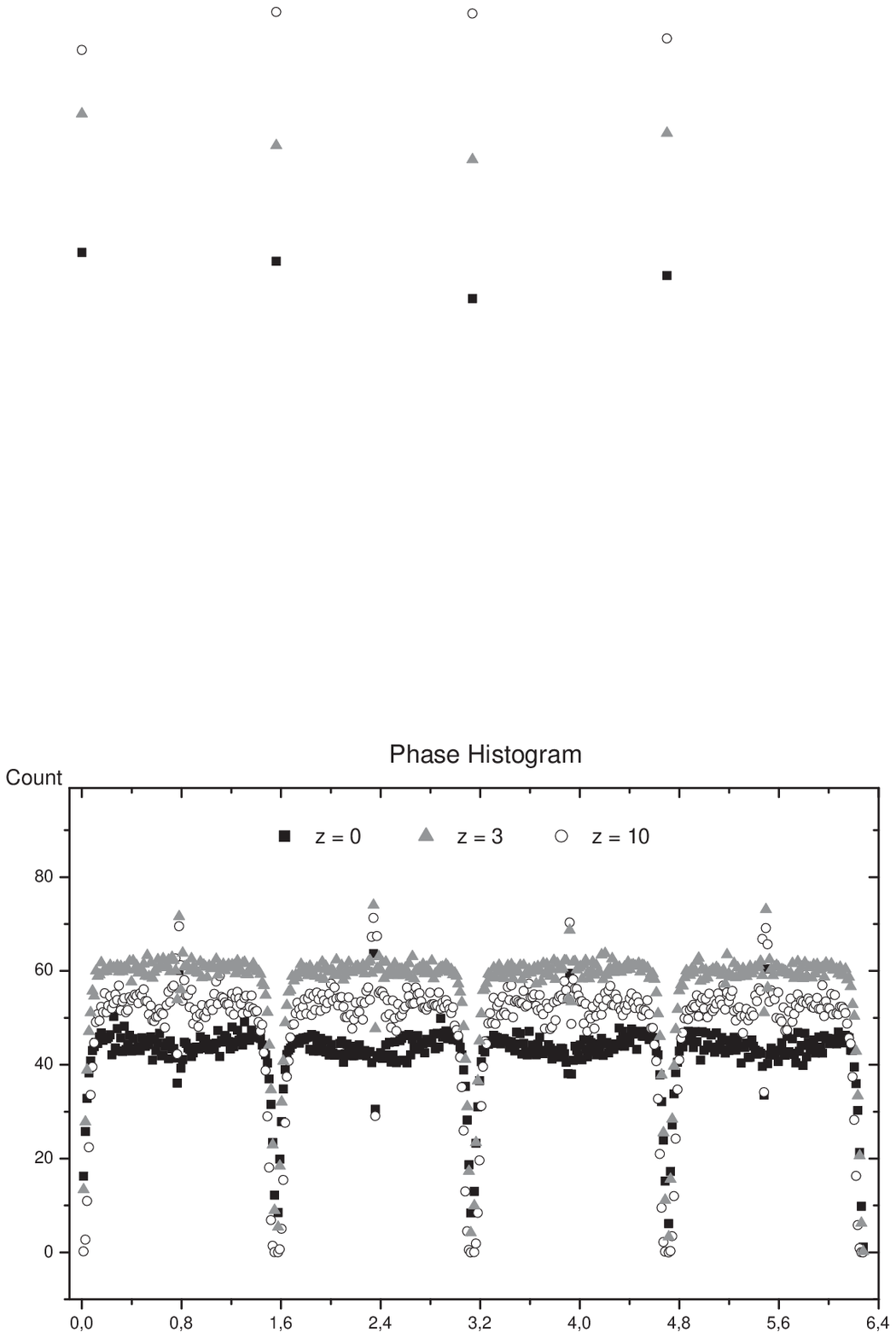}
\caption{\label{fig:epsart} Norms and phases histogram.}
\end{figure}

\begin{figure}
\includegraphics[width=2.8in, height=2.6in]{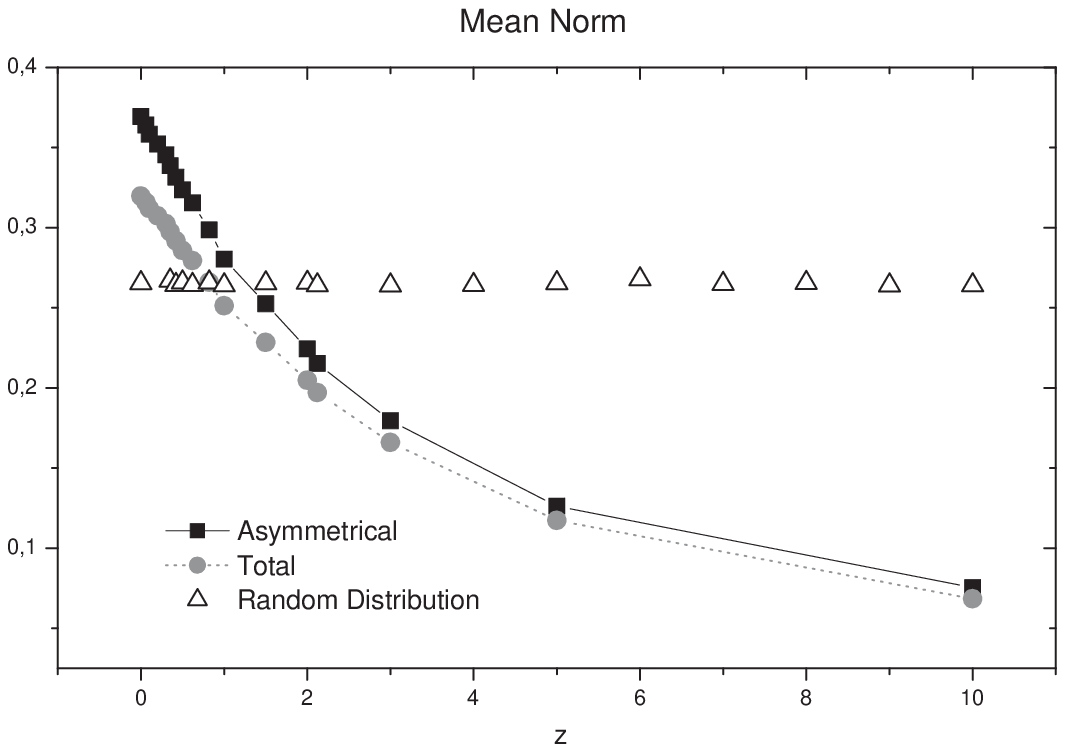}
\includegraphics[width=2.8in, height=2.6in]{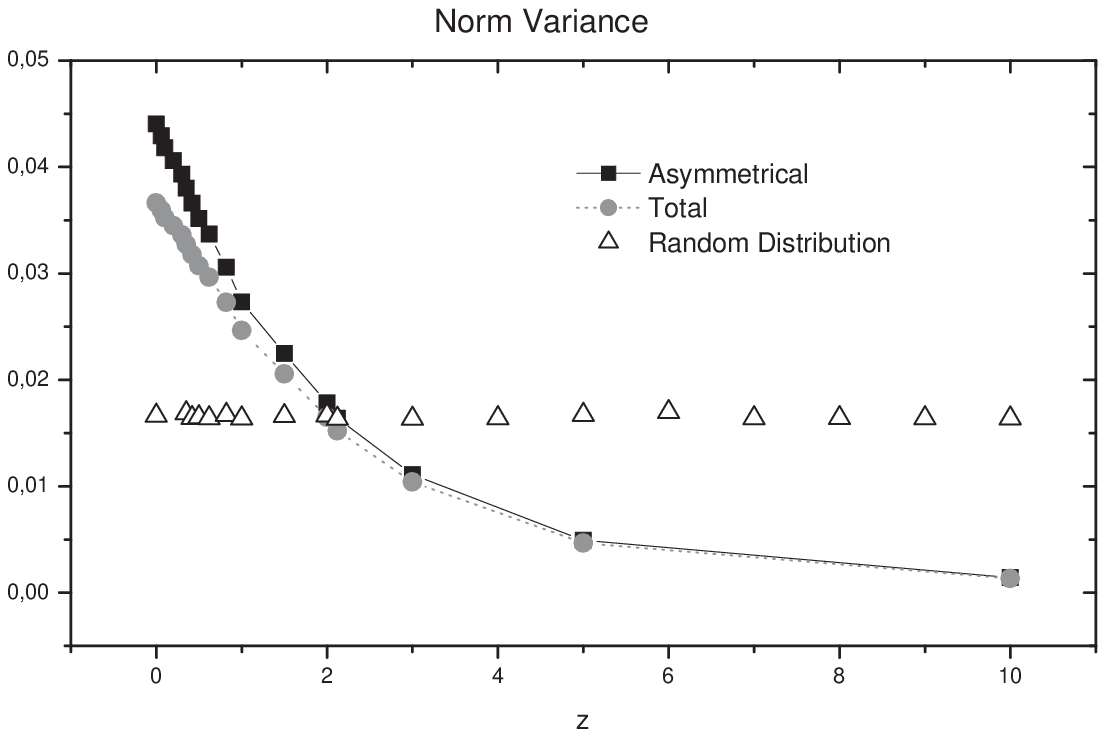}
\includegraphics[width=2.8in, height=2.6in]{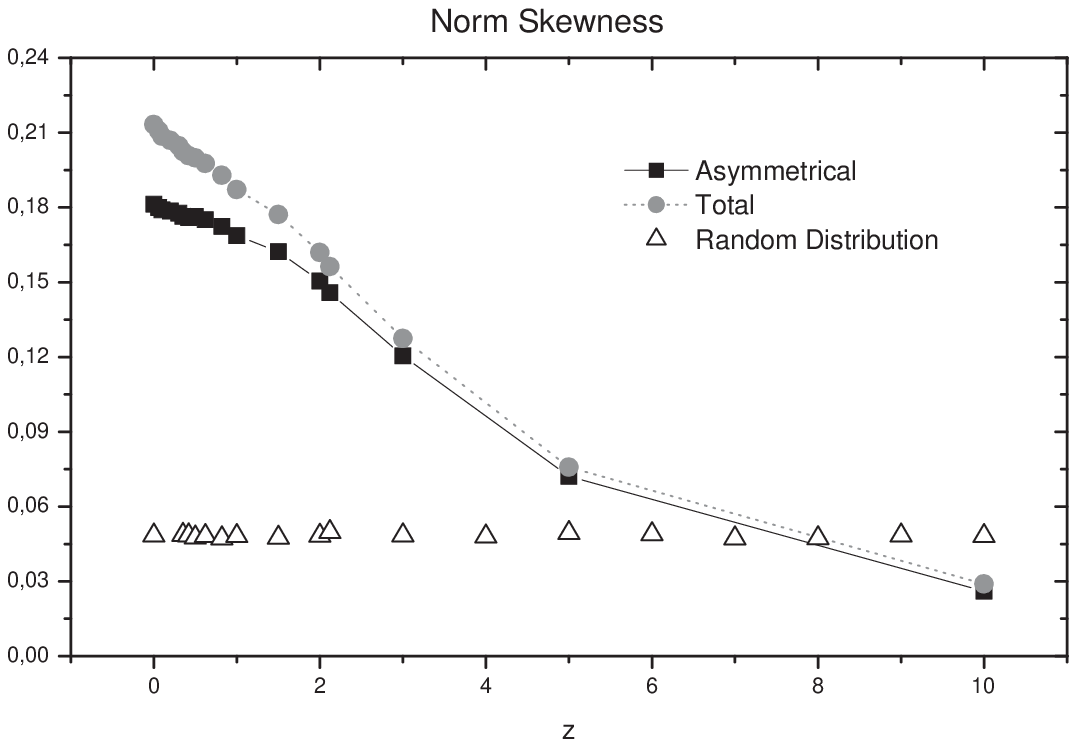}
\includegraphics[width=2.8in, height=2.6in]{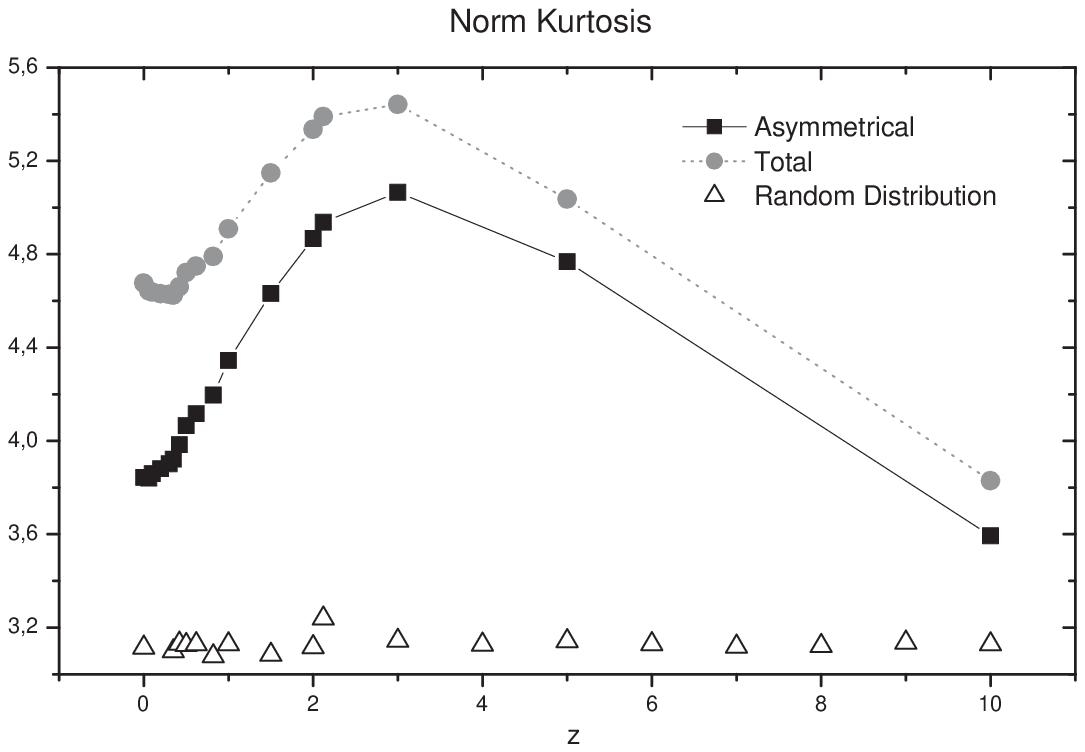}
\caption{\label{fig:epsart} Means, variance, skewness and kurtosis of the
  norms.}
\end{figure}

In Figure 3 we can observe the evolution of the norms and phase 
histogram, as the system evolutes and the structure starts to collapse.
It is evident on the histograms that the statistical distributions of norms is evolving to a 
higher asymmetrical distribution. As indicated by all the statistical moments
for the norms estimated in Figure 4. This behavior is a strong
evidence of an increase of amplitude fluctuations (i.e. inhomogeneity) in the density field.
The distortion that happens on the norm distribution is characterized by an
increase of mean, variance and skewness. However, in the kurtosis parameter,
we can observe that the redshift $z \sim 3$ is a crucial point in the norm distribution, since the
kurtosis start to decrease after this point. At $z \sim 3$, another change in
the norm distribution can be observed, the increase rate in the skewness
parameter is lower after this point.

\begin{figure}
\includegraphics[width=2.8in, height=2.6in]{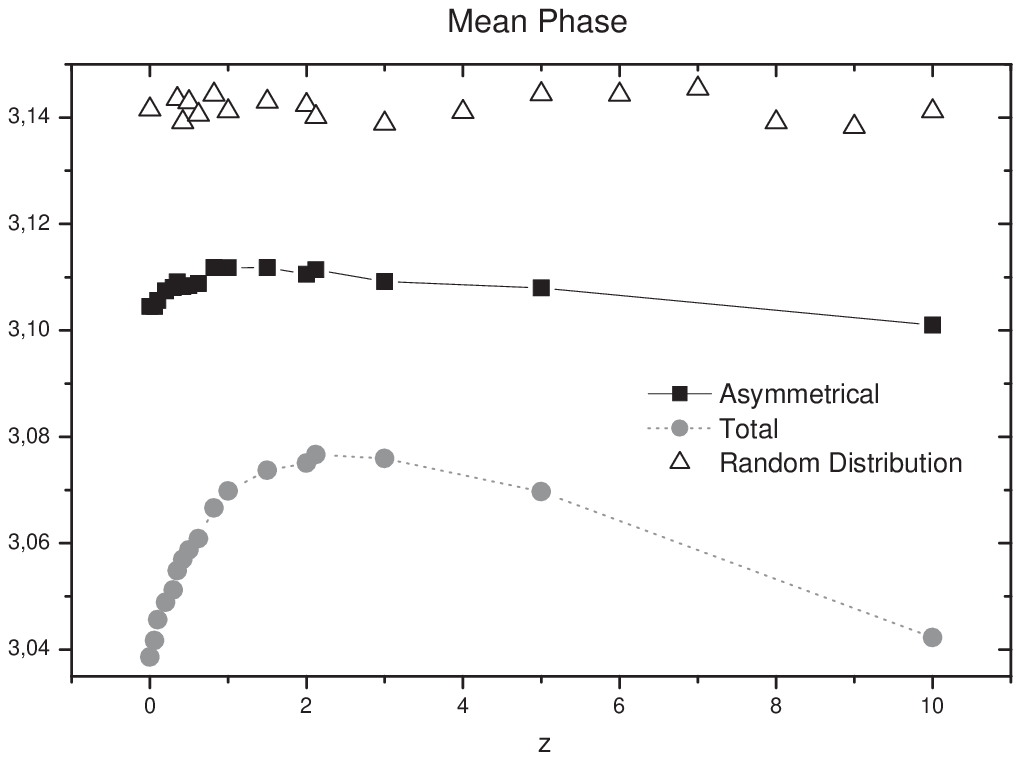}
\includegraphics[width=2.8in, height=2.6in]{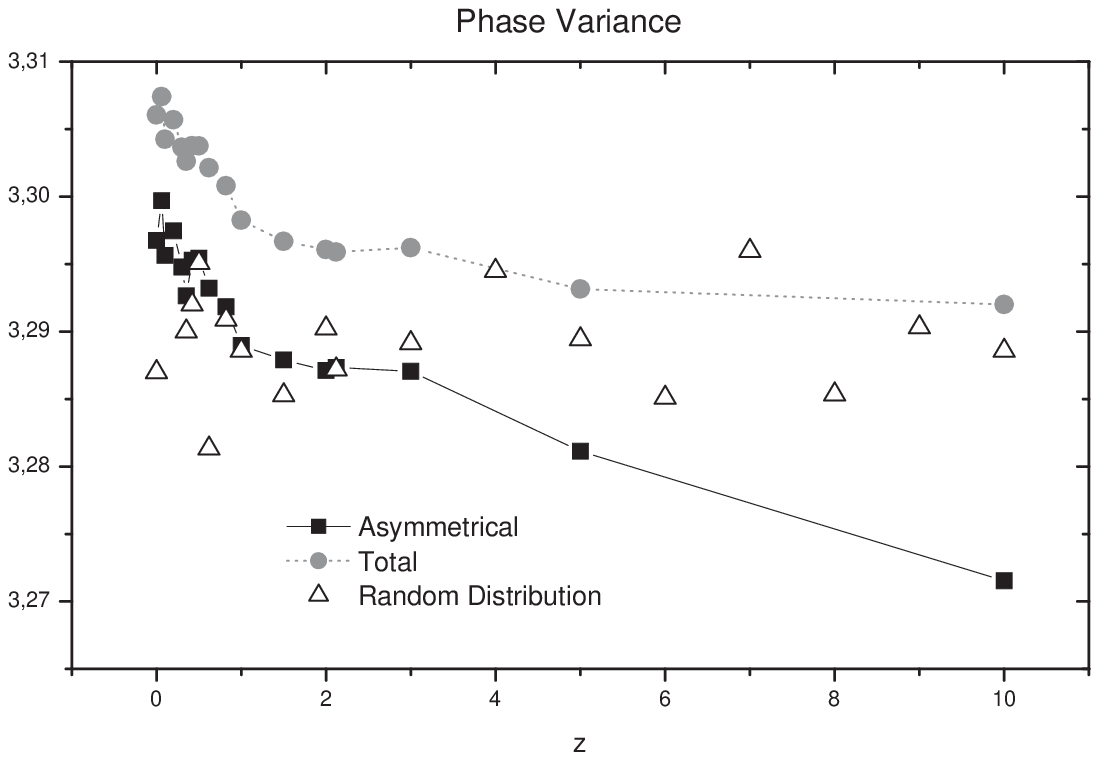}
\includegraphics[width=2.8in, height=2.6in]{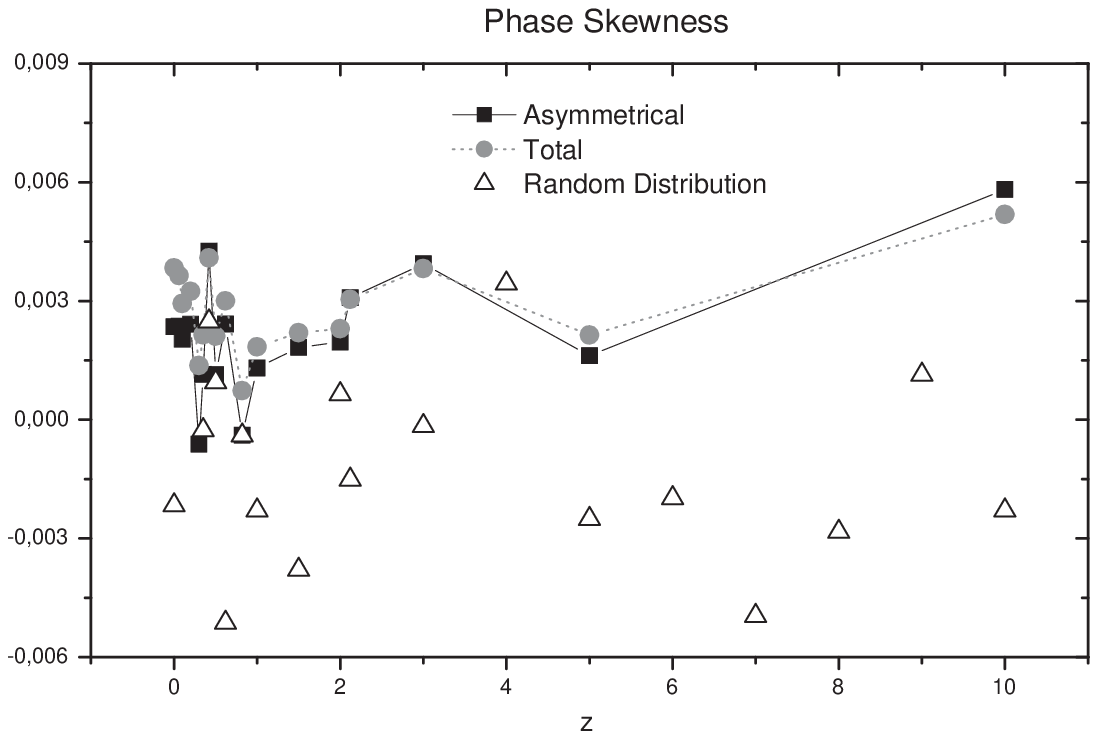}
\includegraphics[width=2.8in, height=2.6in]{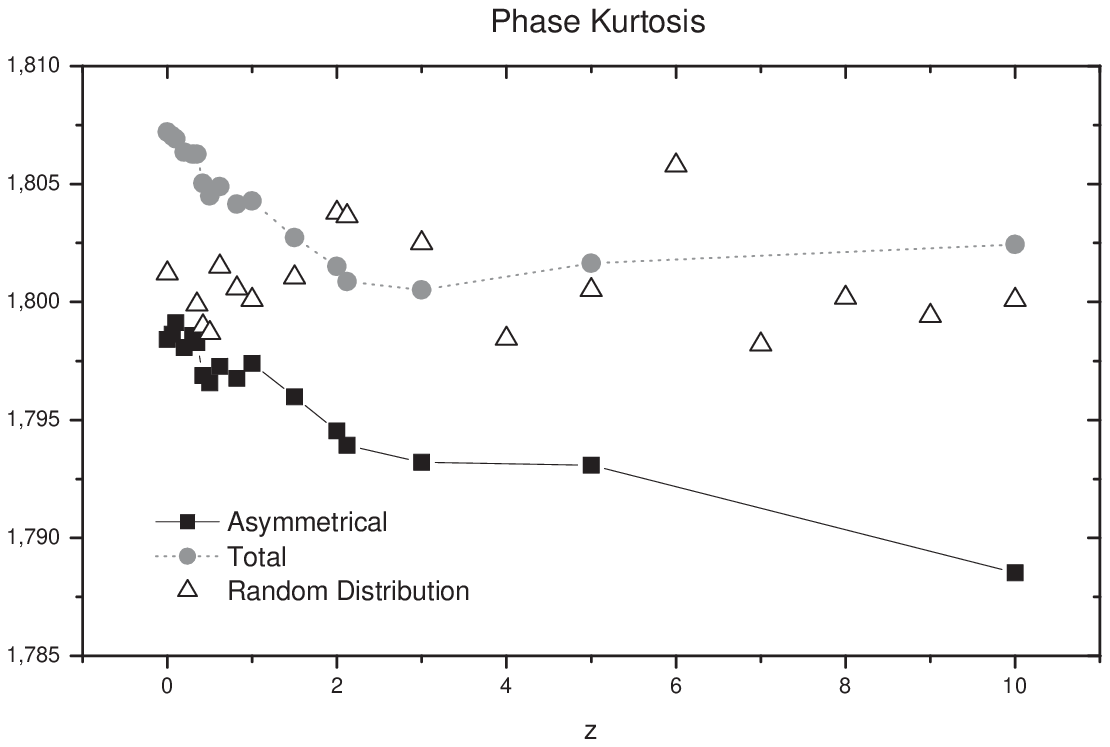}
\caption{\label{fig:epsart} Means, variance, skewness and kurtosis of the
  phases.}
\end{figure}

The distortions in the phase distribution are not so well defined as the case of
the norms, since there are too much statistical fluctuations, as can be
observed in Figure 5. It is difficult to conclude any evidence of anisotropy,
however, the phase distribution 
reveals a higher dispersion and a decrease in the mean phase for redshift lower than $ z \sim 3$.  The
phase variations have small amplitude, but it is possible to infer a considerable
increase in phase variance, and kurtosis, specially for redshift lower than 3, while the mean increases at the beginning 
of the clustering process and start to decline for lower redshifts. Again, the
time scale of $z \sim 3$ seems to play a special hole in the process of
cosmic structure formation.

The interesting behavior of norm and phase distributions for low redshift can be
observed by another feature of the GPA analysis, the number of gradient
vectors in the density field. In figure 6, we show the total number of gradient
vectors and the relative number of asymmetrical vectors. In this
plot we can observe that the total number of vectors is increasing at the
beginning of the clustering process at  $z \leq 10$, has a maximum point at $
z \sim 3$ and start to decline meaningfully for $z < 3$. Not only
the total number of vectors is decreasing, but the relative number of
asymmetrical vectors has a much more stronger decline. At $z \sim 3$ the asymmetrical 
vectors number is more than $80\%$ of the field, while in present time, it is 
nearly $60\%$, less than it was at the beginning of clustering process $\sim 70\%$.

This result is an evidence that, for redshift lower than $z \sim (2-3)$, the local fluctuations in scales of $\sim 0.7
Mpc~h^{-1}$ is getting smoothed. This feature can be explained by the
infalling of matter in dynamically young galaxy clusters in conjunction with
the formation of voids in the surroundings of overdensed regions. This feature
is also indicated by the faster increase of the
amplitude fluctuations (i.e. the mean vectors norm), at the same time, the
underdensed regions are expanding, becoming large voids, where the gradient
established is far below the mean value, or even null.

\begin{figure}
\includegraphics[width=2.8in, height=2.6in]{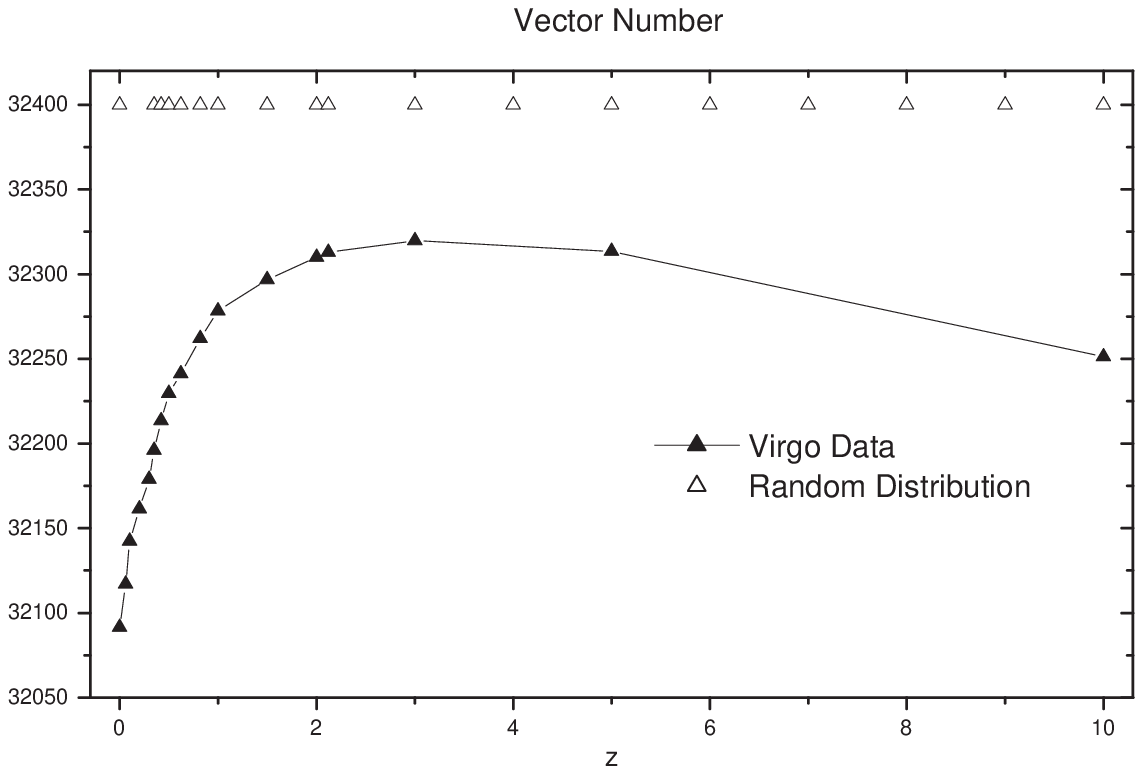}
\includegraphics[width=2.85in, height=2.6in]{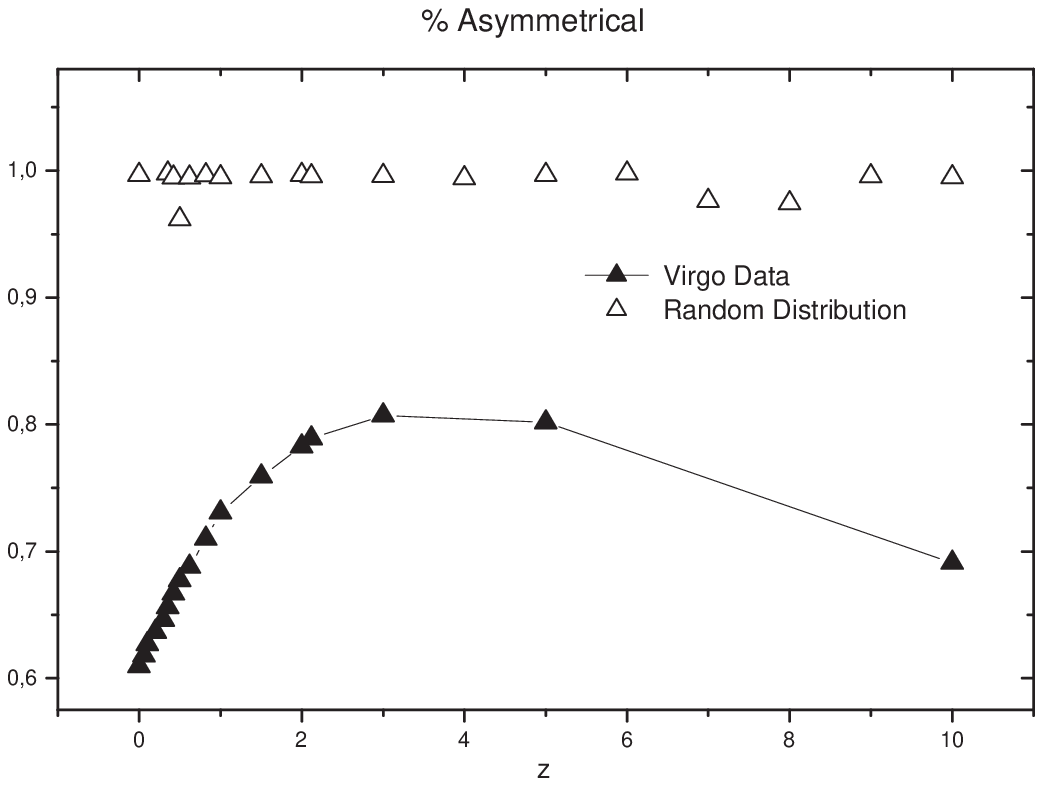}
\caption{\label{fig:epsart} The vectors number and the relative number of
  asymmetrical vectors.}
\end{figure}

Comparing curves in Figures 4 and 5, it is possible to conclude that the field of 
asymmetrical vectors is also a good representation of the behavior of the gradient 
field, since the curves for total and the asymmetrical number of 
vectors present nearly the same behavior. However, the total number of vectors 
is indeed a better statistical assembly. Anyway, the investigation of the
number of the asymmetrical vectors can bring additional information about
symmetry breaking in the evolutionary process. Another point important to
observe is that, as evidenced by the statistical
analysis, the behavior of the pattern evolution of density fluctuations 
field is quite different from a pure random field. Indeed, the effects of time evolution is easily 
differentiable from a random case, specially for lower redshifts.

To better understand the evolution of the gradient pattern for the 
density field, we have performed a nonlinear curve fit to the norm histogram. 
The best fit parameters were obtained for three classes of curves: 
Gaussian, Lorentzian and Log-Normal. The correlation coefficient for some redshift 
values are showed in Table 1 below. In this table we can observe that for a 
redshift $z \sim 10$ the best fit is a Gaussian distribution, which
characterizes a low correlated field. While the field evolves and the
structures start to organize inside clusters after $ z \leq 3 $, the best fit is 
a Log-Normal distribution, which characterizes a higher correlated and
asymmetrical field. This result illustrates the same behavior observed by 
the statistical moment analysis for the gradient field, specially the complex 
correlations illustrated in Figure 7 below.

\begin{table}
\begin{center}
\begin{tabular}{|c|c|c|c|}\hline
z & Gauss & Lorentz & Log-Normal \\\hline\hline
0.0  & 0.45856 & 0.46068 & 0.48097 \\\hline 

0.5  & 0.72328 & 0.72955 & 0.75061 \\\hline

1.0  & 0.85986 & 0.86963 & 0.88910 \\\hline

3.0  & 0.95181 & 0.95854 & 0.96624 \\\hline

5.0  & 0.96835 & 0.96865 & 0.96975 \\\hline

10.0 & 0.98180 & 0.97142 & 0.96782 \\\hline
\end{tabular}
\caption{\small The correlation coefficient for the Gaussian, Lorentz 
and Log-Normal fit over the norm histogram.}
\end{center}
\end{table}

An additional estimate to do with the gradient vectors field is the complex
correlation between pairs of vectors, as defined in section 2.1. This
parameter estimate may be helpful to elucidate how effective are the couplings in the global
gradient field and their temporal relation. Accordingly, we would like
to compare the GPA complex correlation with another correlation factor able to
characterize spatial randomness by analyzing the properties of the field at 
different times, such as the spatial length correlation function, $E_n$ (Cross
\& Hohenberg, 1993) :

\begin{equation}
En(l,t) \equiv \frac{\frac{1}{N}\sum_{i,j=1}^{N}\bar x(t)_n^{(i,j)}\bar x(t)_{n}^{(i+l,j+l)}}{\frac{1}{N} \sum_{i,j=1}^{N}(\bar x(t)_n^{(i,j)})^2},
\end{equation}

\noindent being: $ \bar x(t)_n^{(i,j)} \equiv x(t)_n^{(i,j)}- \left < x \right >_n$,
where $ x(t)_n^{(i,j)}$ is the amplitude matrix in $(i,j)$ position, in a time scale $t$, and $l$ is the
spatial displacement $(l=1,2...)$. The $E_n(l)$ function can be used as a rough quantitative measure of correlations for a
given spatial profile in a fixed time scale, t.

The estimate for the GPA complex correlation and the $E_n(l)$ correlation
function for z=10, 3 and 0 are illustrated in Figure 7. Both
correlation factors represent simple statistic factors, but have completely
different concepts: the former is a global statistic applied in the local fluctuations,
while the second is a straight measure of spatial correlations in the amplitude field. 
The decaying behavior of the spatial correlation function in coupled lattices
presents a typical power-law decay, however, we can observe that the $E_n(l)$
function is unable to predict any temporal relation of the density field, in despite of the GPA complex
correlation that, although it does not concern about spatial correlations in
large scales, it is able to describe the temporal evolution of gradient correlations for the density field in
scales of $0.7 Mpc~h^{-1}$. 

\begin{figure}
\includegraphics[width=2.8in, height=2.7in]{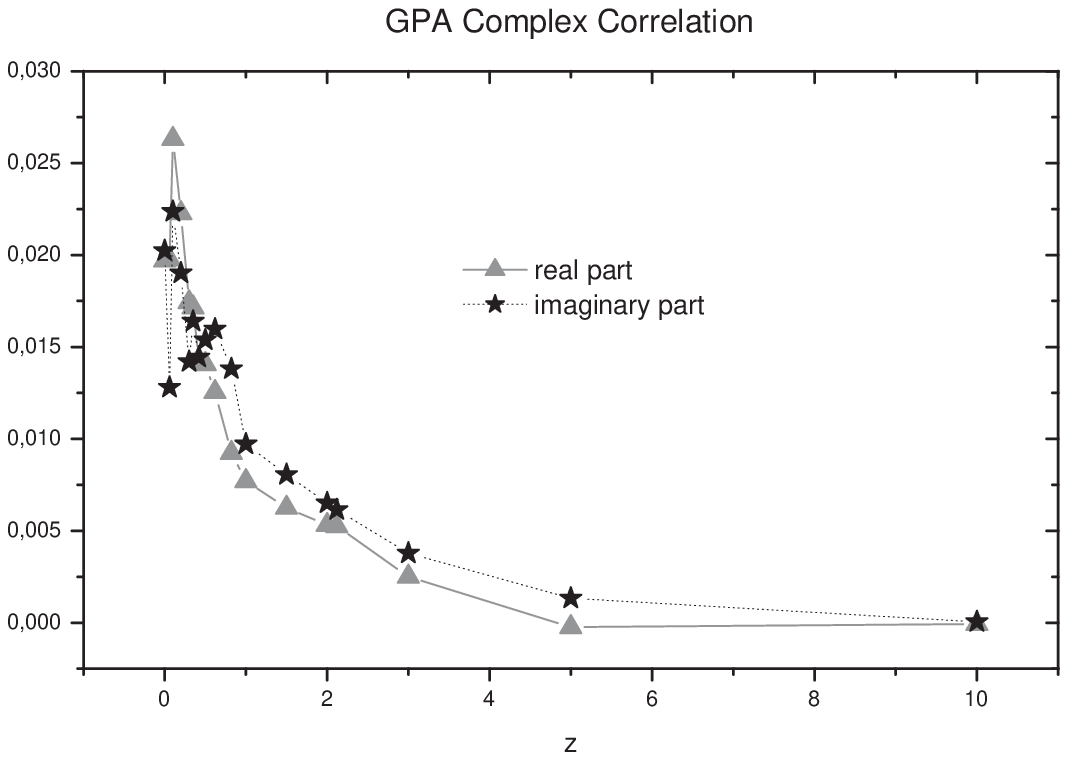}
\includegraphics[width=2.8in, height=2.7in]{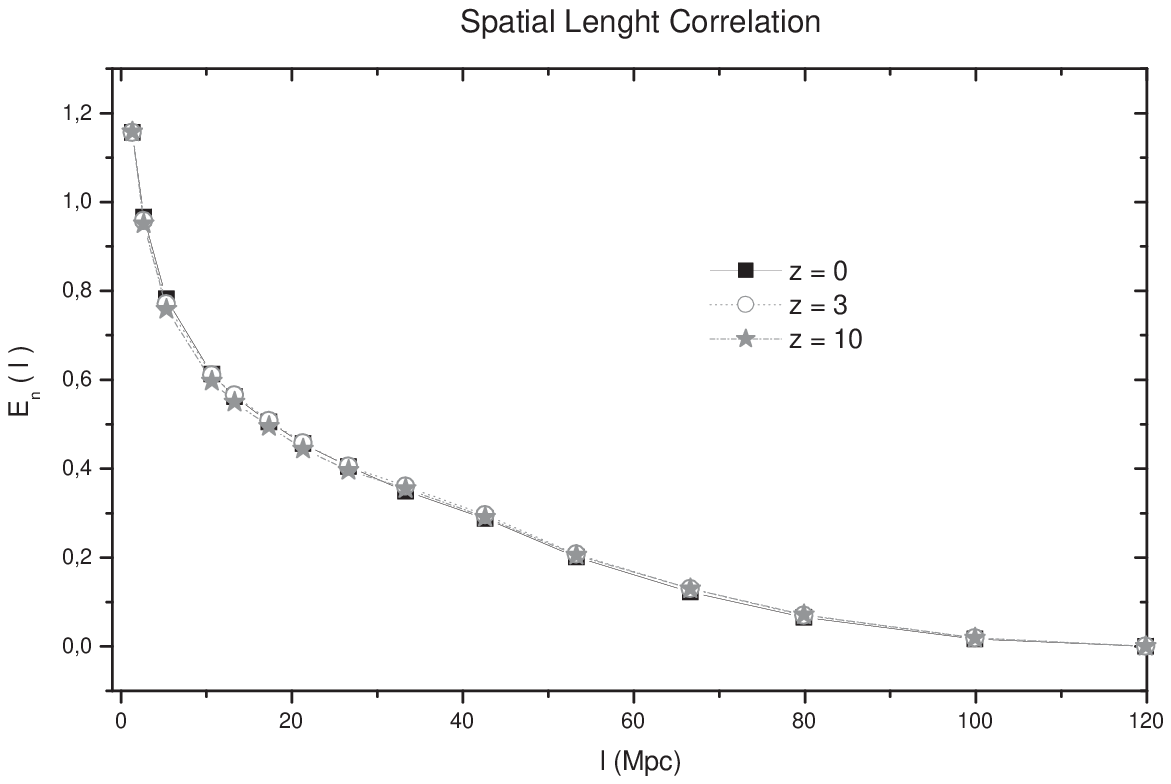}
\caption{\label{fig:epsart} The complex GPA correlation and the spatial length
  correlation.}
\end{figure}

\section{Discussion}
\label{sec:level1}
The late stages of cosmic structure formation are usually modeled by discrete 
numerical simulations, which do not allow us to follow the details of  
the entire gravitational collective phenomena behind the structure 
growth. However, charactering the spatio-temporal evolution of matter 
clustering is an important task in cosmology, since we need to compare 
theoretical predictions with observational data. In this paper we present 
the first test of the gradient pattern analysis as an alternative tool to 
investigate and characterize the nonlinear growth of large scale structure in the 
Universe. In particular, we are checking the ability of the GPA second and 
third gradient moment to describe the pattern evolution in matter clustering. 
 
We have performed the statistical analysis of the norm and phase for  
the gradient vectors estimated over the density fluctuations field for  
some bi-dimensional slices of the Universe. The results show that the  
GPA technique is able to discriminate pattern formation and evolution in  
large scale structures. As could be expected, the mean, variance and skewness of  
norms present an increase distribution distortion as the system evolutes and the  
fluctuations interact in a nonlinear way and the structures start to organize and  
to collapse in clusters and filaments. For the vectors phases, the statistical  
fluctuations are bigger, however the variance and kurtosis are well defined,  
and show, except for the mean values, a sensible distortion in time evolution, as it was expected for the process of  
structure clustering. The distortions in norms and phase distributions can be 
interpreted as an evidence of higher irregularities in the fluctuations  
field for the scale investigated, $0.7 Mpc~h^{-1}$. The GPA complex 
correlation is well defined and corresponds to the increase of amplitude fluctuations and phase  
interactions as the density field evolutes and the statistical behavior starts to  
differentiate from a random field, as observed on the best fit distribution 
over the norms histogram. 

Nevertheless, the most important result of this first GPA investigation on N-body 
cosmological simulations seems to be the {\it turned point} observed in the number of 
asymmetrical vectors in the gradient field for $z \leq 3$. This epoch
probably marks an intensification of galaxy cluster formation with
sub-structures organizing inside clusters. This result suggests the application of the GPA statistical 
analysis for simulations in different scales, in order to have a better
resolution to map substructures in galaxy clusters, as well as large patterns like wall filaments and
voids regions.

Anyway, the decline in the number of gradient vectors may be 
directly related to the expansion of interlocked empty spaces between
structures inside the galaxy clusters or even with the expansion of large empty structures, 
like big voids, which have a spherical expanding feature and can reach sizes up to tens of Mpc 
(Sheth \& Weygaert, 2004). The redshift $z \sim 3$ may represent the time scale of 
the voids structures building, however, to assure this conclusion it is 
necessary a detailed investigation in different scales and resolution.  
We hope that future works with the GPA technique may  
bring some additional information about the size, the spatial correlation and time 
evolution of voids. 
 
In this first test, we may conclude that the statistical analysis of  
the second and third gradient moment may be useful to describe pattern 
evolution in a nonlinear interactive field.  
This result is very stimulating for future investigations of the gradient pattern formation analysis in the  
process of cosmic structure formation. At this moment, we are applying the gradient 
analysis over the DPOSS observational data for galaxy and cluster distribution, 
in order to compare with the temporal evolution characterized for a simulated
\ensuremath\Lambda-CDM Universe. 
The next step on the GPA investigation on cosmic structure formation will be to 
perform the same statistical analysis of the second and third gradient moments 
in different scales, trying to estimate the asymmetrical fragmentation of 
gradient field with the help of the Delaunay triangulation process. We are 
also planning to expand this calculation to a three-dimensional mesh.

{\bf Acknowledgments}\\ \\
Special thanks to {\it Virgo Consortium} team. APAA thanks the 
financial support of CNPq and FAPESB, 
under grant 1431030005400. ALBR thanks the financial support of CNPq, 
under grant 470185/2003-1 and 306843/2004-8.




\end{document}